\documentclass{ws-procs975x65}

\def\ud{\mathrm{d}}

\newcommand{\ket}[1]{|\kern.3ex#1\kern.3ex\rangle}
\newcommand{\bra}[1]{\langle\kern.3ex #1 \kern.3ex|}
\newcommand{\scalar}[2]{\langle\kern.3ex #1 \kern.3ex|\kern.3ex#2\kern.3ex\rangle}
\newcommand{\norm}[1]{\|\kern.3ex#1\kern.3ex \|}

\def\lg{\langle }
\def\rg{\rangle }

\def\ud{\mathrm{d}}

\begin{document}
\title{Physical Hilbert Spaces in Quantum Gravity}

\author{Ma\l kiewicz, Przemys\l aw}

\address{$^1$APC, Univ Paris Diderot, CNRS/IN2P3, CEA/Irfu, Obs de Paris, Sorbonne Paris Cit\'e, France\\
$^2$National Centre for Nuclear Research, Ho\.za 69, Warsaw, Poland\\
E-mail: przemyslaw.malkiewicz@ncbj.gov.pl}

\begin{abstract}
We summarize our investigation of the extent to which the choice of internal clock influences the dynamics in quantum models of gravity. Firstly, at the classical level, we define an extension to the Hamilton-Jacobi theory of contact transformations, which allows for transformations of time coordinates. Secondly, at the quantum level, we employ the extended theory to separate the quantum effects brought by the free choice of internal clock from those originating from inequivalent quantization maps. Next, we show with two examples two kinds of origin of the clock effect in quantum gravitational systems.

\end{abstract}


\bodymatter

\section{Introduction}
Internal clocks in canonical relativity play two distinguishable roles: (i) they restrict the space of physical three-geometries to the ones with a constant value of clock and (ii) they set the simultaneity between those three-geometries belonging to different physical spacetimes (solutions to the Einstein equations). The space of such simultaneous constant-clock physical three-geometries, which may be given by three-metrics and the conjugate momenta, form the so-called reduced phase space.

In the classical formalism, the equations of motion transform suitably upon switching between clocks so that the physics remains unaffected. In the quantum formalism, however, the physical predictions will depend on the choice of clock. This is so because the choice of clock is intimately tied to the definition of reduced phase space and it can be showed that in general reduced phase spaces induced by distinct clocks are not canonically equivalent, see Ref.~\refcite{Haj}. Thus, canonical quantization can not possibly lead to the same results as it establishes unitarily inequivalent theories. The only question, which remains, is about the extent of the physical dissimilarities.

For the spatially homogenous cosmological models, the homogeneity defines the preferred slicing of the spacetimes and clocks are imposed to be constant on these slices. Nevertheless, this still leaves a large freedom in defining internal clock and the corresponding reduced phase space. In the present paper we study homogeneous spacetimes and investigate the ambiguity described in point (ii) and its consequences for the result of quantization. We show that already in this relatively restricted set-up the reduced phase spaces are canonically inequivalent and that this non-equivalence may lead to very different quantum dynamics. 

\section{What is the Hamiltonian constraint?}
From now on we discuss only finite-dimensional phase spaces, which in particular accommodate the spatially homogenous models. The Hamiltonian constraint introduces the constraint surface $S$ in the finite-dimensional kinematical phase space, which is equipped with the symplectic form $\omega$. In $S$ the induced two-form $\omega|_S$ is degenerate and its null direction generates the dynamical orbits. In other words, physical dynamics is induced form $\omega$ by a hypersurface in the kinematical phase space.

All the points in the hupersurface $S$ are physically distinguishable states of the model, because clocks themselves are internal degrees of freedom whose values are meaningful. Therefore, all functions defined in the constraint surface are physical observables. Among them we single out those, which do not vary along the orbits, the so called Dirac observables. The remaining ones are called dynamical observables. The split is essential as in the surface $S$ the Dirac observables form the Poisson algebra, whereas the commutation rule for the dynamical observables is not defined. The Poisson algebra, which extends the algebra of Dirac observables to include dynamical observables can be introduced with the help of an internal clock as follows. A clock is defined as a constraint surface function, which varies monotonically along all the dynamical orbits (for simplicity, let us assume that each value of the clock crosses each orbit once and only once). Next, the induced form $\omega|_S$ is restricted to a fixed value-of-clock submanifold, where it becomes non-degenerate and thus invertible into the Poisson bracket. If the former is repeated smoothly for all values of the clock and all the respective submanifolds, the Poisson bracket is introduced into the whole constraint surface $S$, which gives the commutation rules for all functions on $S$ including dynamical observables. It is clear that different clocks lead to different Poisson brackets. This construction was first introduced in Ref.~\refcite{MCT}.

\begin{figure}[t]
\begin{center}
\includegraphics[width=0.34\textwidth]{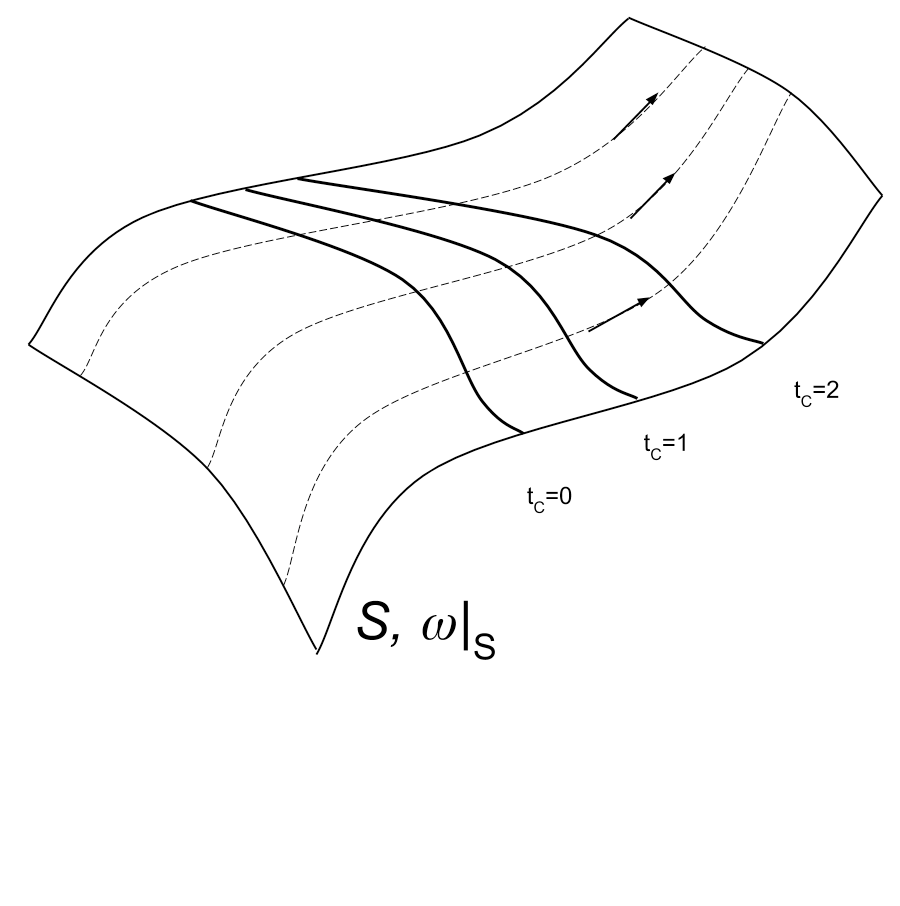}
\includegraphics[width=0.6\textwidth]{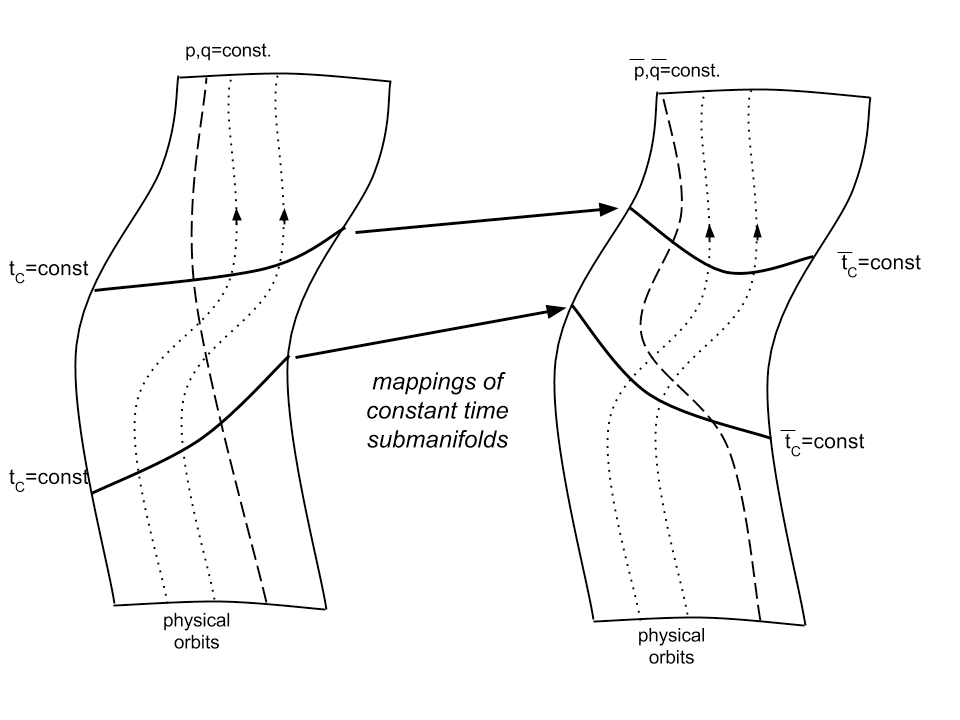}
\end{center}
\caption{Constraint surface $S$ comprises physical orbits, which are integral curves of a field of null vectors of the induced form $\omega|_S$. Further restriction of this form to constant-time manifolds, $\omega|_{S,t_C=const}$, establishes a contact manifold in the constraints surface and a physical Poisson bracket. Different time coordinates induce different physical Poisson brackets, which can be related by constant-time-to-constant-time mappings as illustrated on bottom.} 
\label{F1}
\end{figure}

When a clock is introduced in $S$, we define the reduced phase space as a constant-clock submanifold in $S$. The Cartesian product of the values of clock and the reduced phase space constitutes a contact manifold. The coordinates of the contact manifold $(q^i,p_j)$ are said to be canonical if $\omega|_S$ has the following form:
\begin{equation}\label{con1}
\omega|_S=\ud q^i\ud p_i-\ud t_C\ud H(q^i,p_j,t_C)
\end{equation}
where $t_C$ is a clock. The function $H(q^i,p_j,t_C)$ is called the physical Hamiltonian and it generates the motion via the usual Hamilton equations. It is clear that for another choice of clock, say $\tilde{t}_C$, in general another set of coordinates $(\tilde{q}^i,\tilde{p}_j)$ has to be picked in order to bring $\omega|_S$ to its canonical form:
\begin{equation}\label{con2}
\omega|_S=\ud \tilde{q}^i\ud \tilde{p}_i-\ud \tilde{t}_C\ud \tilde{H}(\tilde{q}^i,\tilde{p}_j,\tilde{t}_C).
\end{equation}

\section{How to relate different reduced phase spaces?}
The Hamilton-Jacobi theory of contact transformations describes passive transformations of canonical coordinates, which preserve the canonical form of $\omega|_S$ of eq. (\ref{con1}) in a fixed time coordinate, $t_C$. For more details, see e.g. Ref.~\refcite{AbMa}. Therefore, it is too limited to be used for relating different clock-based reduced phase spaces, and it needs to be extended. The extension turns out to be quite natural once one grasps the relation between the choice of clock and the Poisson bracket. Firstly, one allows for passive transformations, which preserve the canonical form of $\omega|_S$ as in eq. (\ref{con2}) upon changing clock coordinate to $\tilde{t}_C$. Because the clock is transformed in this case, such transformations can no longer be regarded as canonical. However, because they preserve the canonical form of $\omega|_S$, the Hamilton equations of the physical motion can be formed wrt $\tilde{t}_C$, and we may discuss transformations leading to these equations. They are realized via a family of symplectomorphisms between constant-clock submanifolds in $S$ defined with use of different clocks. Let us fix a function $t_C\mapsto \tilde{t}_C(t_C)$. Then, the transformation from a constant-clock submanifold $\mathcal{M}_{t_C}$ to a constant-clock submanifold $\mathcal{M}_{\tilde{t}_C(t_C)}$ such that all the transformed points remain in the same orbits is a symplectomorphism. If for both clocks there is a respective set of canonical coordinates, then the symplectomorphisms are defined as maps preserving the canonical form of the respective symplectic forms, that is, $$\ud q^i\ud p_i=\omega|_{S,~\mathcal{M}_{t_C}}\mapsto
~\omega|_{S,~\mathcal{M}_{\tilde{t}_C(t_C)}}=\ud\tilde{q}^i\ud\tilde{p}_i\, .$$ 
As a result, we obtain a family of transformations, which in Ref.~\refcite{MCT} we call formally canonical transformations.

The application of the extension of the H-J theory is two-fold. Firstly, having a canonical description of any mechanical system, we can obtain another canonical formulation with respect to any time coordinate. The way to do it is to postulate a specific form of a family of formal canonical transformations and then to determine the new canonical coordinates from the requirement that the maps preserve the orbits\footnote{This leads to a set of algebraic equations, which may be found in Ref.~\refcite{CQ}}. The second application, which is relevant for our purpose, is to determine the clock effect brought by quantization. Specifically, if a family of formally canonical transformations becomes a family of formally unitary operators upon quantization, then any discrepancies in the corresponding quantum theories are due to distinct clocks rather than distinct quantizations. In what follows we explain why.

The generalized transformations, which include clock's transformation, and which rely on respective families of formally canonical mappings, are active transformations as they move points in the surface $S$ along the orbits (see Fig. \ref{F1}). Therefore, in general, they transform an initial physical observable into another one. The only exceptions are the Dirac observables as they do not vary along the orbits. Therefore, the corresponding families of unitary operators ensure true, physical (not only formal) unitary correspondence between the quantized Dirac observables. This removes quantization ambiguity in the discussed comparison. At the same time, when the families of unitary mappings are used to relate evolving observables, they assign the same (or, unitarily equivalent) operators to physically different dynamical observables. For this reason, we expect that a fixed dynamical observable will, in general, be represented by inequivalent operators, when viewed with different clocks. This was indeed shown to be the case in Ref.~\refcite{Kas}.

\section{Quantum averaging}
The first source of discrepancies, which appear in semiclassical considerations, comes from computations of expectations values of a dynamical observable at fixed values of a respective clock. Assume a phase space with canonical coordinates $(x,p)\in\mathbb{R}^2$, which are the Dirac observables and thus there is no Hamiltonian in this phase space. Furthermore, there is a clock, $t_C\in\mathbb{R}$, and thus dynamical observables are functions $\mathcal{O}(x,p,t_C)$. Now, as we transform the clock $t_C\mapsto \tilde{t}_C=t_C+x^k$, the dynamical observables become in the new clock coordinate $\mathcal{O}(x,p,\tilde{t}_C-x^k)$ (let $k$ be an unspecified integer). Let us quantize the phase space,
$$
x\mapsto\hat{x}\psi(x):=x\psi(x),~~p\mapsto\hat{p}\psi(x):=-i\frac{\ud}{\ud x}\psi(x),~~\psi\in{L}^2(\mathbb{R},\ud x),
$$
and pick a normalized state from the Hilbert space, say, $
\psi(x)=\pi^{-\frac{1}{4}}\exp\left(-\frac{x^2}{2}\right)
$. It is clear that the expectation value of any Dirac observable, $\lg\psi|\hat{\mathcal{O}}(x,p)|\psi\rg$, is independent of the choice of clock, so let us focus on dynamical observables. To simplify the subsequent consideration as much as possible, let us consider the observable $\mathcal{O}=t_C^2$, which in the other clock reads $\mathcal{O}=(\tilde{t}_C-x^k)^2$. Next, we compute expectation values of $\hat{\mathcal{O}}$ with respect to different clocks. We find the respective minimal values of $\lg\psi|\hat{\mathcal{O}}|\psi\rg$ with respect to $t_C$ and $\tilde{t}_C$, $$
\textrm{min}\lg\psi|\hat{\mathcal{O}}|\psi\rg\big|_{t_C}=0 \textrm{~~and~~}
\textrm{min}\lg\psi|\hat{\mathcal{O}}|\psi\rg\big|_{\tilde{t}_C}=\frac{(2k)!}{2^kk!}\,,$$ where $k$ is odd. We conclude that the minima of some dynamical observables can be altered as much as one wishes by picking sufficiently large values of $k$. Above we have considered the simplest possible setup to clarify this source of discrepancies. In fact, they must arise in any setup and in a more complex one they may arise in a much less straightforward way. For more details on this example, see Ref.~\refcite{MCT}.

\section{Quantum corrections}
The other source of physical dissimilarities has to do with dynamics, which picks up quantum corrections. These corrections are essential in quantum gravity models, which resolve spacetime singularities. When quantization is performed with different clocks, the corrective terms will have to be different. It is so because the corrections themselves are dynamical observables (they modify dynamics), and any transformation of clock modifies dynamical observables. This renders the correction variant with the transformations. A nice example is provided by a flat FRW model, which is formally identical to a particle freely moving in the half-line. The phase space is given by $(q,p)\in\mathbb{R}_+\times\mathbb{R}$. The dynamics of the particle terminates at the origin point of the half-line $q=0$, however, quantization may resolve this singularity. For instance, the so called coherent state quantization  introduces a repulsive potential, which prevents the particle from reaching the origin point as shown in Ref.~\refcite{FRW}. At the semiclassical level, the classical Hamiltonian, $H=p^2$, is replaced by the semiclassical one, $\check{H}=p^2+\frac{K}{q^2}$, where $K$ is a constant. As we will show, the interpretation of the extra term $\propto q^{-2}$ largely depends on the clock used to derive it. Consider clock transformations of type: $t_C\mapsto \tilde{t}_C=t_C+\mathcal{D}(q,p)$, where $\mathcal{D}$ we shall call `delay function'. We find in Ref.~\refcite{DQG} that
$$
\omega|_S=\ud q\ud p-\ud t_C\ud p^2=\ud (q+2p\mathcal{D})\ud p-\ud \tilde{t}_C\ud p^2
$$
and that the pair $(\tilde{q}=q+2p\mathcal{D},\tilde{p}=p)$ is canonical with respect to clock $\tilde{t}_C$. It can be shown that the family of formally canonical mappings in this case reads $
(q,p)\big|_{t_C}\mapsto (\tilde{q},\tilde{p})\big|_{\tilde{t}_C}
$. Any quantizations, which are related by a corresponding formally unitary map, would assign the same operators to $q$ as to $\tilde{q}$, and to $p$ as to $\tilde{p}$, as well as to any respective compound observable. Also, the derivations of respective semiclassical dynamics would follow in parallel. Consequently, the semiclassical Hamiltonian in the new clock reads $\check{H}=\tilde{p}^2+\frac{K}{\tilde{q}^2}$, which pulled-back to the original canonical coordinates reads $\check{H}=p^2+ \frac{K}{(q+2p\mathcal{D})^2}
$. Thus, the comparison can be made in fixed coordinates $(q,p)$ and, as contour-plots of semiclassical Hamiltonians in $(q,p)$ in figure 2 show, the quantum dynamical corrections are indeed tied to the clock involved in the quantization.

\begin{figure}[t]
\centering
\includegraphics[width=0.32\textwidth]{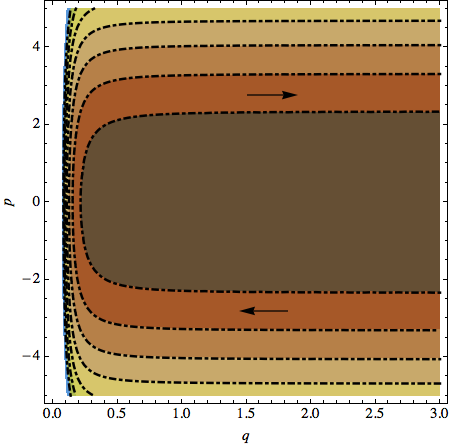}
\includegraphics[width=0.32\textwidth]{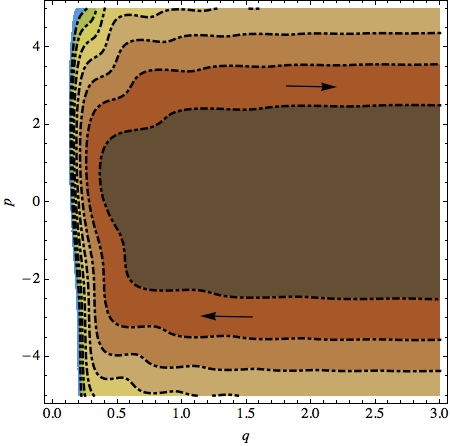}
\includegraphics[width=0.32\textwidth]{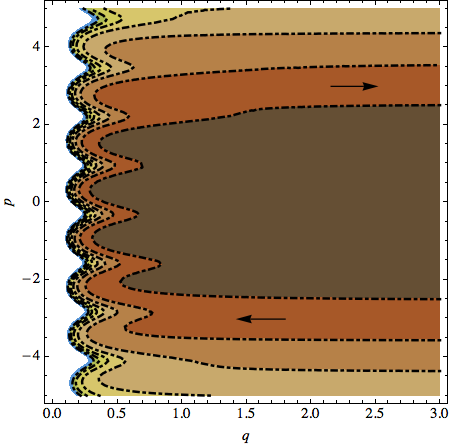}
\label{F2}
\caption{Contour-plots of three semiclassical Hamiltonians derived with various clocks and compared in the fixed coordinates $(q,p)$. They represent some semiclassical features of respective quantum dynamics of a particle in the vicinity of $q=0$, where the particle rebounds. They show strong dependence of dynamics on the clock involved in quantization process.}
\end{figure}

\section{Conclusions}
The resolution of the problem of dynamics in quantum gravity is of fundamental significance. We found that a quantum theory of a gravitational model strongly depends on the clock involved in the quantization process. Thus, quantum dynamics of the gravitational field escapes the usual interpretation. Many further questions can be addressed within the theory of clock transformations presented herein, which may help to make a progress towards the consistent interpretation of quantum gravity.

\section*{Acknowledgements}
This work was supported by Narodowe Centrum Nauki with decision No. DEC-2013/09/D/ST2/03714.

\end{document}